\documentclass{elsart1p}
\usepackage{graphicx}
\usepackage{amssymb}
\begin{document}
\begin{frontmatter}
%
%
%
\title{Role of axial-vector mesons near the\\
chiral phase transition}
%
%
\author[munich]{Chihiro Sasaki},
\author[nagoya]{Masayasu Harada}
and
\author[munich]{Wolfram Weise}
\address[munich]{Physik-Department,
Technische Universit\"{a}t M\"{u}nchen,
D-85747 Garching, Germany}
\address[nagoya]{Department of Physics, 
Nagoya University,
Nagoya, 464-8602, Japan}
\begin{abstract}
We present a systematic study of the vector--axial-vector mixing 
(V-A mixing) in the current correlation functions and its evolution 
with temperature within an effective field theory.
The $a_1$-$\rho$-$\pi$ coupling vanishes at the critical temperature
$T_c$ and thus the V-A mixing also vanishes.
A remarkable observation is that even for finite $m_\pi$ 
the $\rho$ and $a_1$ meson masses are almost degenerate at $T_c$. 
The vanishing V-A mixing at $T_c$ stays approximately intact.
\end{abstract}
\end{frontmatter}
%

In the presence of hot matter 
the vector and axial-vector current correlators are mixed due 
to pions in the heat bath. At low temperatures this process is 
described in a model-independent way in terms of a low-energy 
theorem based on chiral symmetry and consequently the vector spectral
function is modified by axial-vector mesons through the 
mixing theorem~\cite{vamix}. 
The validity of the theorem is, however, limited to temperatures 
$T \ll 2f_\pi$, where $f_\pi$ is the pion decay constant in vacuum.
At higher temperatures one needs in-medium correlators 
systematically involving hadronic excitations other than pions. 
In this contribution
we show the effects of the mixing (hereafter V-A mixing), and how the 
axial-vector mesons affect the vector spectral function near the chiral 
phase transition, within an effective field theory~\cite{our}.


A model based on the generalized hidden local symmetry (GHLS) 
describes a system including the axial-vector meson explicitly, 
in addition to the pion and the vector meson, consistently with the 
chiral symmetry of QCD~\cite{ghls:tree}.
We will use the GHLS Lagrangian as a reliable basis which 
describes the spectral function sum rules~\cite{ghls}.

The critical temperature $T_c$ for the restoration of chiral 
symmetry in its Wigner-Weyl realization is defined as the temperature 
at which the vector and axial-vector current correlators, $G_V$ and
$G_A$, coincide and their spectra become degenerate. Thus, chiral 
symmetry restoration implies $\delta G = G_A - G_V = 0$ at $T_c$.
Let us consider $\delta G$ changing with temperature intrinsically.
To achieve $\delta G = 0$ at the critical temperature, 
we assume non-dropping $\rho$ mass at $T_c$ and adopt the 
following ansatz of the temperature dependence of the {\it bare} 
axial-vector meson mass, $M_{a_1}^2(T) = M_\rho^2 + \delta M^2(T)$:
\begin{equation}
\delta M^2(T) = \delta M^2(T=0)\Theta(T_f-T)
{}+ \delta M^2(T=0)\Theta(T-T_f)\frac{T_c^2-T^2}{T_c^2-T_f^2}\,,
\end{equation}
where we schematically introduce the ``flash temperature'' 
which controls how the mesons experience partial restoration of 
chiral symmetry. The values of $T_c$ and $T_f$ are taken in a 
reasonable range as indicated, for example, by the onset of the 
chiral crossover transition observed in lattice QCD~\cite{LGT}:
$T_c = 200$ MeV and $T_f = 140$ MeV.

\begin{figure}
\begin{center}
\includegraphics[width=6.5cm]{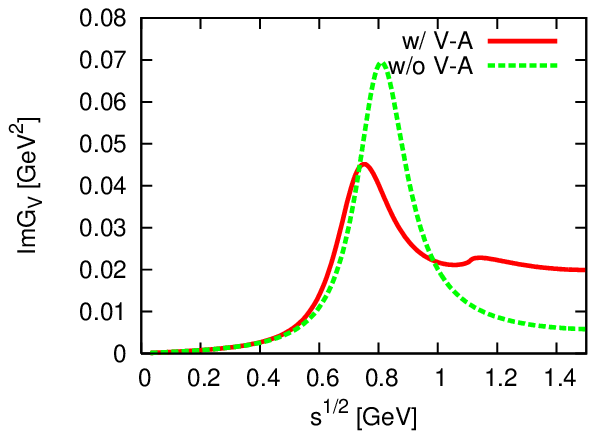}
\includegraphics[width=6.5cm]{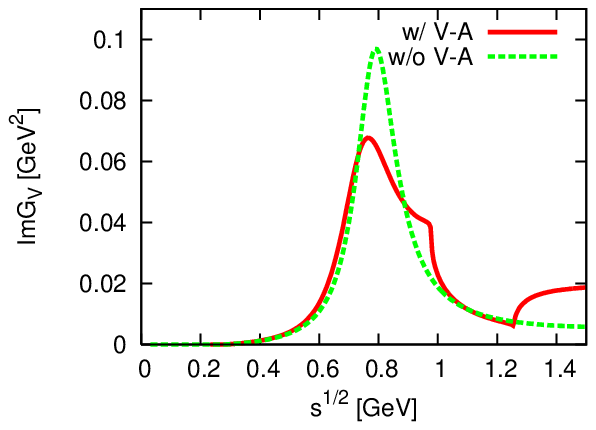}
\caption{
The vector spectral function 
at $T/T_c = 0.8$ calculated in the $\rho$-meson rest frame.
The left side figure is calculated for $m_\pi = 0$
and the right side for $m_\pi = 140$ MeV.
The solid curve is obtained in the full calculation. The dashed 
line is calculated eliminating the axial-vector meson and hence
V-A mixing from the theory.
}
\label{vamix}
\end{center}
\end{figure}

Figure~\ref{vamix} (left) shows the vector spectral function
in the chiral limit.
Two cases are compared; one includes the V-A mixing and the other
does not. The spectral function has a peak at $M_\rho$ and a 
broad bump around $M_{a_1}$ due to the mixing. The height of the
spectrum at $M_\rho$ is enhanced and a contribution above 
$\sim 1$ GeV is gone when one omits the $a_1$ in the calculation.
A difference between the two curves becomes more significant
above $T_f$ where partial restoration of chiral symmetry sets in.
For finite $m_\pi$ the energy of the time-like virtual $\rho$ meson
splits into two branches corresponding to the processes, 
$\rho + \pi \to a_1$ and 
$\rho \to a_1 + \pi$, with thresholds $\sqrt{s} = M_{a_1} - m_\pi$
and $\sqrt{s} = M_{a_1} + m_\pi$.
This results in
the threshold effects seen as a shoulder at $\sqrt{s}=M_{a_1}- m_\pi$
and a bump above $\sqrt{s}=M_{a_1} + m_\pi$ 
in Fig.~\ref{vamix} (right).
Note that the enhancement of the spectrum for $m_\pi \neq 0$
is due to the change of the phase space factor $(s - 4m_\pi^2)^{3/2}$.

\begin{figure}
\begin{center}
\includegraphics[width=6.5cm]{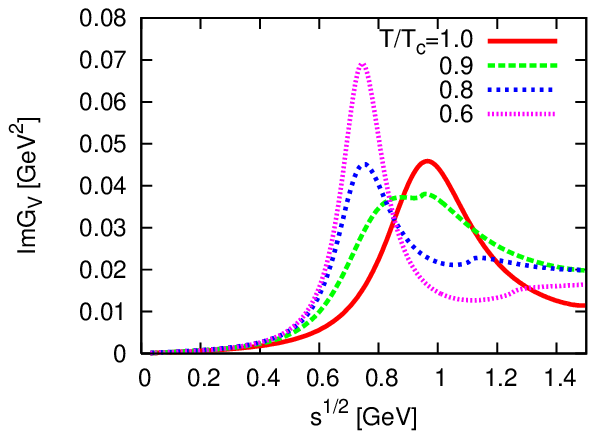}
\includegraphics[width=6.5cm]{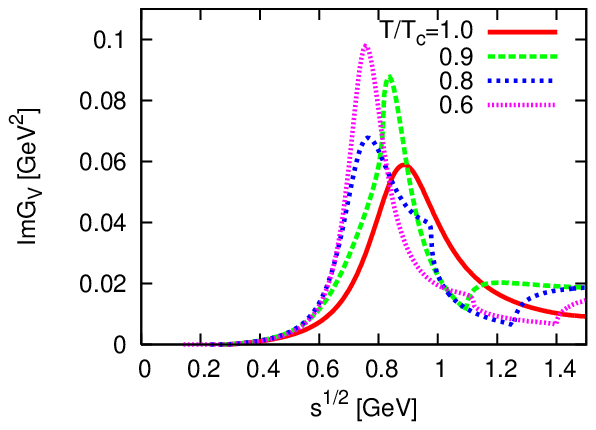}
\caption{
The vector spectral function
for $m_\pi = 0$ (left) and
for $m_\pi = 140$ MeV (right) at several temperatures 
$T/T_c = 0.6$-$1.0$.
}
\label{tdep}
\end{center}
\end{figure}

Figure~\ref{tdep} (left) shows the temperature dependence of the
vector spectral function in the chiral limit.
One observes a systematic downward shift of the $a_1$ enhancement
with increasing temperature, 
while the peak position corresponding to the $\rho$ pole mass
moves upward.
At $T/T_c = 0.9$ the two bumps begin to overlap:
the lower one corresponds to the $\rho$ pole, and the upper one
to the $a_1$-$\pi$ contribution.
Finally at $T=T_c$, $M_{a_1}$ becomes degenerate 
with $M_\rho$ around $\sqrt{s} \simeq1\,$GeV
and the two bumps are on top of each other. 
The V-A mixing eventually vanishes there.
This feature is a direct consequence of vanishing coupling of $a_1$ 
to $\rho$-$\pi$.
Figure~\ref{tdep} (right) shows the effect of finite pion mass
in the vector spectrum. Below $T_c$ one observes
the previously mentioned threshold effects moving downward with 
increasing temperature. It is remarkable
that at $T_c$ the spectrum shows almost no traces of 
$a_1$-$\rho$-$\pi$ threshold effects.
This
indicates that at $T_c$ {\it the $a_1$ meson mass nearly equals
the $\rho$ meson mass and the $a_1$-$\rho$-$\pi$ coupling
almost vanishes even in the presence of explicit
chiral symmetry breaking.}


In summary,
we have presented a detailed study of V-A mixing in the current
correlation functions and its evolution with temperature.
In the chiral limit the axial-vector meson contributes significantly 
to the vector spectral function; the presence of the $a_1$ 
reduces the vector spectrum around $M_\rho$
and enhances it around $M_{a_1}$.
For physical pion mass $m_\pi$, the $a_1$ contribution 
above $\sqrt{s} \sim M_{a_1}$ still survives although the bump is 
somewhat reduced.
When assuming both dropping $\rho$ and $a_1$ masses, 
the major change is
a systematic downward shift of the vector spectrum~\cite{our}.

Studying dilepton production in relativistic heavy-ion collisions
is an interesting application.
The present investigation may be of some relevance for the high
temperature and low baryon density scenarios encountered
at RHIC and LHC.

\subsection*{Acknowledgments}

The work has been supported in part by BMBF and by the DFG cluster 
of excellence ``Origin and Structure of the Universe''.

%
%
%

%
\end{document}